\documentclass[aps,prl,twocolumn,showpacs,groupedaddress]{revtex4}
\usepackage{amsmath,amsfonts,graphicx,color,float}

\begin{document}
\floatplacement{figure}{H}

\title{Current driven rotating kink mode in a plasma column with a non-line-tied free end}

\author{I. Furno$^1$, T.P. Intrator$^1$, D.D. Ryutov$^2$, S. Abbate$^1$, T. Madziwa-Nussinov$^1$, A. Light$^1$, L. Dorf$^1$, and G. Lapenta$^1$}

\affiliation{ $^{1}$ Los Alamos National Laboratory, M.S. E526,
Los Alamos, NM 87545, USA\\
$^{2}$ Lawrence Livermore National Laboratory, Livermore CA 94551}

\date{\today}

\begin{abstract}
{First experimental measurements are presented for the kink
instability in a linear plasma column which is insulated from an
axial boundary by finite sheath resistivity. Instability threshold
below the classical Kruskal-Shafranov threshold, axially
asymmetric mode structure and rotation are observed. These are
accurately reproduced by a recent kink theory, which includes
axial plasma flow and one end of the plasma column that is free to
move due to a non-line-tied boundary condition.}
\end{abstract}

\maketitle

The current driven kink instability is a magnetohydrodynamic (MHD)
instability which affects current carrying plasmas in Nature and
laboratory. The kink mode structure and stability condition are
strongly dependent on the system geometry and the boundary
conditions (BCs). Kruskal\cite{Kruskal:1958} and Shafranov
\cite{Shafranov:1956} (hereafter referred to as KS) considered
first the ideal MHD stability of a cylindrical plasma column with
magnetic field components ($0, B_{\theta}, B_z$) using cylindrical
coordinates ($r, \theta, z$). For an infinitely long (equivalent
to periodic axial BCs) column, they obtained a linearly unstable
helical kink mode of structure $\mathbf{\xi} = e^{i(\theta + 2 \pi
z/L)}$ when the plasma current $I_p$ exceeds the Kruskal-Shafranov
limit
\begin{equation}
I_{KS}=(2\pi)^2 a^2 B_z / (\mu_0 L) \label{Kruskal-Shafranov.eq}
\end{equation}
where $a$ and $L$ are, respectively, the radius and length of the
current channel, and $\xi$ is the displacement of the plasma
column from the equilibrium position. The KS theory has been quite
successful in predicting the behavior of toroidal plasmas for
which the periodic BCs yield a proper accounting for the finite
length of the system. In linear systems, however, substantial
deviations from KS predictions can result from different axial
BCs.

The importance of the BCs has long been
recognized\cite{Solove'ev:1971, Raadu:1972} and is of particular
relevance to the stability of line-tied flux ropes in space
physics ({\em c.f.} Ref.~\cite{Baty:1997} and survey Ref.~\cite{Hood:1992}),
and astrophysical jets\cite{Meier:2001}. In
recent years, there has been a renewed interest in the stability
of a line-tied plasma column in laboratory devices (see Refs.
\cite{Lanskii:1990, Ryutov:2004,Hegna:2004} and references
therein). The kink stability of a plasma column with line-tied
ends has been investigated in linear devices, where
line-tying is attributed to the presence of highly conducting end
plates\cite{Bergerson:2006}, and in open systems to a local
discontinuity for the Alfv\'{e}n velocity that forms a virtual
boundary around the system\cite{Hsu:2004,Zuin:2004}.

In this Letter, we experimentally investigate the external kink
instability in conditions where one end of the plasma column is
line-tied to the plasma source, and the other end is not line-tied
and therefore free to slide over the surface of the end-plate. The
latter BC is a result of plasma sheath resistance that insulates,
at least partially, the plasma from the end-plate. Compared to the
line-tied case, we find significant differences in the kink mode
structure and lower critical current for the onset of the kink.
The axial flow velocity, the direction of the axial magnetic
field, and the rotation frequency of the kink mode are all
correlated. The experimental results agree with the predictions
from a recent theory of the external kink instability for a
slender plasma column\cite{Ryutov:2005}.

\begin{figure}[!t]
\includegraphics[scale=1]{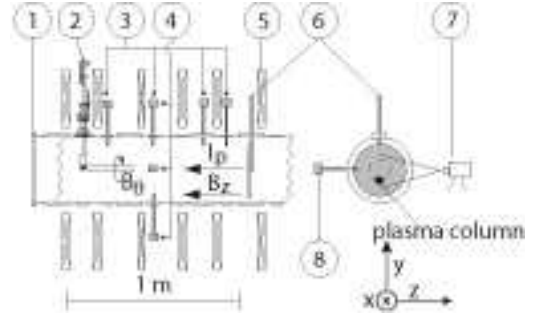}
\caption{Experimental setup with diagnostics and coordinate system.
(1)  Vacuum vessel; (2) plasma gun; magnetic probes in (3) an axial
array and (4) an azimuthal array; (5) external coils; (6) external
anode. On the right, an axial cut near the external anode. The fast
camera (7) is located at the midplane and views the plasma column
along the $x$ direction. The bi-dimensional magnetic probe (8)
measures ($\delta B_x, \delta B_y$) at the edge. Also schematically
shown is the plasma column whose end rotates at the external anode.}
\label{setup.fig}
\end{figure}
The experiments are conducted in the Reconnection Scaling
Experiment (RSX) which uses electrostatic plasma guns to generate
the plasma\cite{Furno:2003}. Figure \ref{setup.fig} shows the
experimental setup with a view of the plasma gun located at $z=0$
and radially inserted into the center of the RSX cylindrical
vacuum vessel ($4$ m length and radius $r_{wall} = 0.2$ m). A
hydrogen arc plasma is ejected to form a cylindrical plasma column
in a constant, uniform, axial magnetic field of $B_z =0.012$ T
generated by external coils, Fig. \ref{setup.fig}-(5). Electron
density and temperature have Gaussian profiles with half-maximum
radius $r_0 \approx 2$ cm such that $r_0 \ll r_{wall}$. Central
electron density and temperature are $n_{e0} = 1-3 \times 10^{19}$
m$^{-3}$ and $T_{e0}=5-14$ eV, and axially decrease away from the
gun. The axial velocity $v_z \approx 3-5 \times 10^{4}$ m$^{-1}$ of the
plasma has been estimated by solving the momentum balance
equations for dominantly axial flow constrained by experimentally
measured axial gradients in pressure and density. A current is
driven in the plasma by negatively biasing the gun with respect to
an external anode ($0.05$ m$^2$ stainless steel plate) which is
electrically isolated from the vacuum vessel. The external anode
can be positioned at different axial locations $z=0.3~-3 $ m
allowing the length $L$ of the current-carrying plasma column to
be varied.

The MHD activity is monitored by two arrays of magnetic probes
that include a total of seven coils inserted in the vacuum vessel
at $r_p=0.15$ m to measure the azimuthal magnetic field $B_\theta$.
In the axial array, Fig. \ref{setup.fig}-(3), four magnetic probes
are positioned at $\theta = \pi /2 $ (top of the vessel) and
axially located at $z = 0.14, 0.48, 0.62$ and $0.76$ m. In the
azimuthal array, Fig. \ref{setup.fig}-(4), the four magnetic
probes are located at $z=0.48$ m and azimuthally equispaced by $\pi
/ 2$. One magnetic probe is shared by the two arrays.
Signals are digitized at $20$ MHz.

\begin{figure}[!t]
\includegraphics{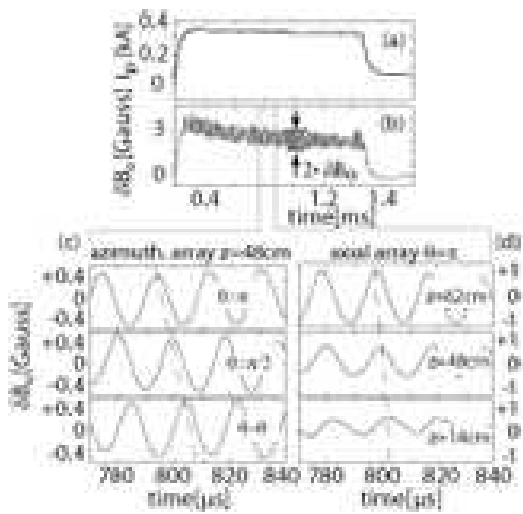}
\caption{Time histories of (a) the plasma current; (b) azimuthal
magnetic field $B_{\theta}$ at ($z=0.48$ m, $\theta = \pi / 2 $, $r
= 0.15$ m). Expanded view of the temporal evolution of $\delta
B_{\theta}$ from azimuthal (c) and axial (d) arrays of magnetic
probes during the stationary phase of the discharge.}
\label{rot_prop.fig}
\end{figure}

In Fig. \ref{rot_prop.fig}(a), the time evolution of the plasma
current measured at the external anode is shown for a discharge
with $L=0.92$ m. The plasma current starts at $t=0$ when the bias
is turned on (arc discharge starts at $t=-1 $ ms) and after
$\approx 150 $ $\mu$s reaches a stationary phase which lasts for
$\approx 1.4$ ms. The total plasma current $I_p \approx 320$ ~A
during the stationary phase is determined by the bias voltage and
can be varied in the range $I_p = 0.05 -1$ kA. Figure
\ref{rot_prop.fig}(b) shows the time evolution of $B_{\theta}$ as
measured by the magnetic probe located at $z=0.48$ m, $\theta =
\pi/2$.

For an expanded time window during the stationary phase,
measurements of the perturbed azimuthal magnetic field $\delta
B_\theta $ from the azimuthal, Fig. \ref{rot_prop.fig}(c), and the
axial, Fig. \ref{rot_prop.fig}(d), arrays of probes show
oscillations with $m=1$ azimuthal periodicity at the frequency $f
\approx 50$ kHz. These oscillations are observed when $I_{p}$
exceeds a current threshold $I_{crit}$ (see below), but are not
detectable for $I_{p} < I_{crit}$.
The observed instability growth time $\tau_{G} \approx 4$ $\mu$s is of
the order of the axial Alfv\'{e}n time.

\begin{figure}[!t]
\includegraphics{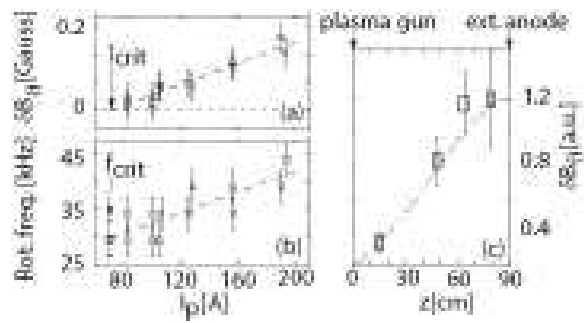}
\caption{(a) Amplitude of the perturbed azimuthal magnetic field,
$\delta B_{\theta}$, and (b) rotation frequency of the mode during
the stationary phase as a function of $I_p$. (c)
Axial structure of $\delta B_{\theta}$ close to the instability
threshold for $I_p / I_{crit} \approx 1.2$ showing non line-tied BC
at the external anode. The dot-dashed line shows the structure of
$\delta B_{\theta}$ as predicted from theory.}
\label{mode_structure.fig}
\end{figure}

The $m=1$ azimuthal structure is consistent with a plasma column
rotating as a rigid body at frequency $f$. The phase of the
signals from the axial array increases linearly along the $z$
direction, Fig. \ref{rot_prop.fig}(d), indicating a rotating
helical current channel. The kinked deformation is a right(left)
handed helix if $J_z \cdot B_z > 0$ $(< 0)$, as expected from a
paramagnetic kink\cite{Hsu:2004}. The direction of rotation
reverses when reversing the external magnetic field $B_z$. In both
cases, the mode rotation is such that the helix always {\em
screws} into the external anode. Both the rotation frequency and
the average amplitude of the $\delta B_\theta $ oscillations at
the dominant frequency during the stationary phase scale linearly
with $I_p$ as shown in Fig. \ref{mode_structure.fig}(a,b) where
the plasma current is varied in the range $86$ A $< I_p < 190$ A
for a plasma column length $L = 0. 92$ m. The plasma current
corresponding to the limit $\delta B_{\theta} \rightarrow 0$
corresponds to the kink stability threshold which for these data
is $I_{crit} = 70 \pm 7$ A. The rotation frequency at the stability
threshold is $f_{crit} = 28 \pm 3.5$ kHz, Fig.
\ref{mode_structure.fig}(b).

\begin{figure}[!t]
\includegraphics{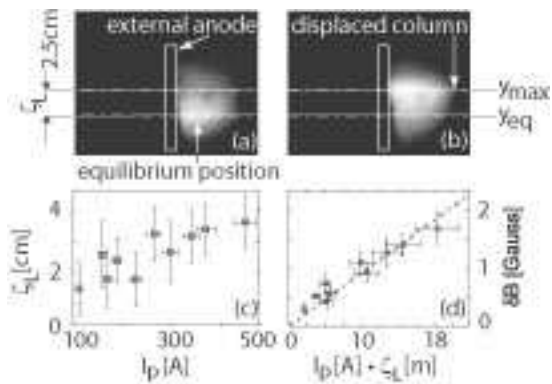}
\caption{Optical and magnetic measurements near the external anode
demonstrate PNLT BC. As the plasma gun shoots from right to left,
a fast gated image of visible emission shows (a) the equilibrium,
$y_{eq}$, and (b) the displaced, $y_{max}$, position of the plasma
during the rotation. (c) Measured optical displacement $\zeta_L =
| y_{max} -y_{eq} |$ for different $I_{p}$ values. (d) Magnetic
field perturbation $\delta B = (\delta B_x ^2 + \delta B_y
^2)^{1/2}$ versus $I_p \zeta _L$ compared with predictions from
Eq. (\ref{non-linetying.eq}) (dashed line). $\zeta _L$ and $\delta
B_{x,y}$ are measured $0.01$ m in front of the external anode.}
\label{non_line_tying.fig}
\end{figure}

Figure \ref{mode_structure.fig}(c) shows the axial structure of the $m
=1 $ mode close to criticality for $I_p \approx 1.2 \times
I_{crit}$. While a line-tied BC holds at the gun, the large
amplitude of the mode at $z \ge 0.76$ m can only occur if the
plasma slides over the external anode and is therefore not
line-tied. This is confirmed by images of the plasma
column at the external anode as shown in Fig.
\ref{non_line_tying.fig}. $H_{\alpha}$ emission is
imaged with a Cooke DiCam Pro intensified camera\cite{Hemsing:2005}
that views the external anode edge perpendicularly to the $z$ axis, Fig. \ref{setup.fig}-(7).
In Figs. \ref{non_line_tying.fig}(a-b), two fast-gated ($200$
ns gate $\ll$ Alfv\'{e}n time) images show
respectively the plasma column in the equilibrium position and at
the maximum displacement $\zeta _L$ during the
rotation.

The observed sliding of the plasma column over the external anode
surface is not consistent with a line-tied BC, which might be
presumed at the external anode for time scales less then a
$100-200$ $\mu$s resistive diffusion time. Deviations from
line-tying may appear if the plasma is insulated from the external
anode by a finite plasma sheath resistance. The degree of
insulation is measured by the ratio $\kappa$ of the Alfv\'{e}n
transit to inductive decay times of the current\cite{Ryutov:2005}.
We estimate $\kappa \approx 15$ from the formula $\kappa \approx
(c_s/v_z) [c / (a \omega_{pi}) ]^2 \beta_e^{1/2}$ where $c_s =
(T_e / m_i)^{1/2}$, $\beta_e = n_e T_e/B_z^2$ and $\omega_{pi}$ is
the plasma frequency (derivation in Ref. \cite{Ryutov:2005}) for
typical plasma parameters $a \approx 2$ cm, $T_e \approx 10$ eV,
$n_e \approx 10^{19}$ m$^{-3}$, $v_z / c_s \approx 1$. For the
present experiments, we conclude that $\kappa \gg 1$ and the
plasma is insulated from the conducting boundary by finite sheath
resistance. Ryutov and coauthors \cite{Ryutov:2005} have shown
that for $\kappa \gg 1$ the BC can be formulated as $(\partial
\zeta /
\partial z + i (k_0/2) \zeta )|_{z=L} = 0 $ where $k_0 =
B_{\theta} / (a B_z) = \mu_0 I_p /(2 \pi a^2 B_z)$ with Cartesian
complex displacement $ \zeta = \xi_x(z,t) +i \xi_y(z,t)$. In the
following, we will refer to this condition as to {\em perfect
non-line-tying} (PNLT) BC and we will show that it holds at the
external anode, after using an equivalent condition that can be
directly compared with experimental data.

With a thin plasma column approximation ($a \ll r_{wall}, L$) and
assuming that $r=0$ is the equilibrium position, the Cartesian
components of the perturbed magnetic field at $r = r_p \gg a$ can
be expressed as $\delta B_{x,y} = (a^2 / r_p^2) B_z (\pm \partial
\xi_{x,y} / \partial z + k_0 \xi_{y,x}$). Using the PNLT
condition, in the vicinity of the external anode the perturbed
magnetic field can be expressed as
\begin{equation}
\delta B = (\delta B_x ^2 + \delta B_y ^2)^{1/2} = I_p \zeta _L (3
\mu_0) / (4 \pi r_p ^2 ) \label{non-linetying.eq} \end{equation}
where $\zeta _L = [ \xi_x(L,t)^2 + \xi_y(L,t)^2 ]^{1/2}$.

We use magnetic measurements combined with images
near the external anode to test Eq.~(\ref{non-linetying.eq}),
equivalent to the PNLT condition. Images such as those in
Fig.~\ref{non_line_tying.fig}(a,b) determine the maximum
displacement $y_{max}$ and equilibrium $y_{eq}$ positions of
emissivity along a line parallel to the external anode
surface and axially spaced $0.01$ m in front of it. The image in
Fig.~\ref{non_line_tying.fig}(b) is synchronized to a magnetic probe
at the same axial location but azimuthally separated by $\pi /2 $
to provide the position of maximum column displacement.
In Fig.~\ref{non_line_tying.fig}(c),
the displacement $\zeta _L = |y_{max} - y_{eq}|$ is shown for a
series of discharges with 100 A $< I_p <
500$ A. In Fig.~\ref{non_line_tying.fig}(d), the magnetic
perturbations ($\delta B_x$, $\delta B_y$) are measured at $r_p =
0.17$ m and $1$ cm in front of the external anode using a
bi-dimensional magnetic probe. In Fig.~\ref{non_line_tying.fig}(d),
the experimental data show excellent agreement with the PNLT predicted scaling in Eq.
(\ref{non-linetying.eq}) (shown as a dashed line). We therefore
conclude that the PNLT BC applies to the external anode.

For a PNLT BC at $z = L$ and with axial plasma flow velocity
$v_z$, the eigensolution for the MHD equation of motion (Eq. (57)
in Ref. \cite{Ryutov:2005}) is $\zeta = C ( e^{i k^{+}z} - e^{i
k^{-}z} ) e^{ i \theta -i \omega t} $ with axial wave numbers
$k_{\pm}=\pm (\pi /2 L) [ 1 \mp (1 - M ^2)^{1/2}]$ where $M = v_z/
( \sqrt{2} \bar{v}_A)$ is the Alfv\'{e}n Mach number, and
$\bar{v}_A$ is calculated using the average density $\bar{n}_e =
(2 / a^2) \int_0^{a} n_e(r) r dr$. In Fig.
\ref{mode_structure.fig}(c), the axial structure of $\delta B_{
\theta}$ from experimental data for $I_p \approx 1.2 \times
I_{crit}$ (squares) and from the theoretical eigensolution
(dot-dashed line) are compared. The good agreement suggests a
robustness of the eigensolution for slightly super-critical
regimes. The total axial phase shift at criticality (not shown
here) is also consistent within the experimental uncertainties
with the expected eigensolution for Mach numbers in the range $M
=0.25 - 0.4$.

The critical current in this case is
\begin{equation}
I_{crit} = \frac{1}{2} \frac{ (2 \pi a) ^2 B_z}{\mu_0 L}(1 -
M^2)^{1/2} \equiv \frac{I_{KS}}{2} (1 - M^2)^{1/2}
\label{marginal_stability.eq}
\end{equation}
showing that the kink mode is unstable at half the KS current for
vanishing flow and smaller currents when flow exists. The
eigenfrequency $\omega$ at the criticality has a finite real
component
\begin{equation}
Re(\omega) = - \frac{\sqrt{2} \pi {\bar{v}}_A}{L} M (1 -M
^2)^{1/2}.
\label{frequency_marginal_stability.eq}
\end{equation}
The perturbed plasma column rotates at frequency $Re(\omega) = 2
\pi f$, constant along the column. The rotation is driven by axial
plasma flow from the gun to the external anode along the
helically-perturbed column as well as a Doppler shifted $k_{\pm}
\cdot v_z$, such that the helix always {\em screws} into the
external anode, consistently with the experimental observations.

\begin{figure}[!t]
\includegraphics{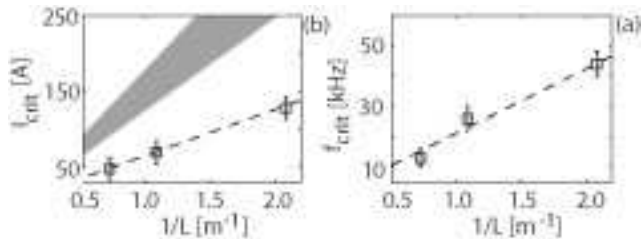}
\caption{For three different column lengths ($L=0.48, 0.92, 1.38$
m): (a) critical current $I_{crit}$ (b) rotation frequency at the
criticality $ f_{crit}$ as a function of the inverse column length
$1/L$. The grey region indicates predictions from KS theory.
Dotted lines show linear fits.} \label{scaling_L.fig}
\end{figure}

To investigate the applicability of Eqs.
(\ref{marginal_stability.eq}) -
(\ref{frequency_marginal_stability.eq}), figure
\ref{scaling_L.fig} shows (a) the critical current and (b) the
rotation frequency at the criticality as a function of the inverse
column length $1/L$ which is in agreement with a predicted linear
scaling.

\begin{figure}[!t]
\includegraphics{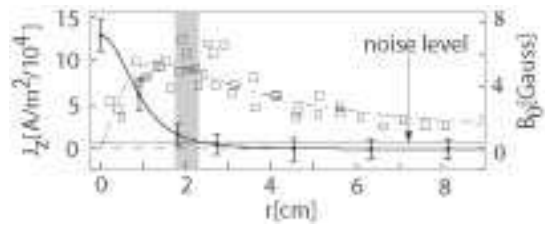}
\caption{Radial profile of current density (solid line) from
measured azimuthal magnetic field (squares) at the criticality.
The vertical grey region indicates the error bounded estimate of $a$
at the edge of the current channel.}
\label{current_density.fig}
\end{figure}

The system comprising Eqs. (\ref{marginal_stability.eq}) -(
\ref{frequency_marginal_stability.eq}) provides a theoretical
relationship between $I_{crit}$, $f_{crit}$, $a$, $M$ and the
average plasma density (or Alfv\'{e}n speed). This was solved
using experimental measurements of the electron density profile at
criticality together with deduced scalings $I_{crit} = 59 / L$ [A]
and $ f_{crit} = 2.4 \times 10^4 /L $ [Hz] from the data in Fig.
\ref{scaling_L.fig}. We obtained a radius $a = 2 \pm 0.2$ cm which
is in agreement with the experiment (see Fig.
\ref{current_density.fig}) and Mach number $M = 0.25 - 0.4$. The
corresponding axial flow velocity $v_z \approx 3-4.8 \times 10^4
$ m$^{-1}$ is also consistent with our independent estimate. In Fig.
\ref{scaling_L.fig}, predictions from KS theory using the computed
$a$ are compared to the experimental data.

For the external kink, the radius $a$ is expected at the position
where conductivity, magnetic diffusion time, and axial
current density $J_z(r)$ are negligibly small.
This is tested using profile measurements at the
criticality. Figure \ref{current_density.fig} shows a Bennet
profile least-squares fit of $J_z(r)$
at the instability threshold, as obtained from azimuthal magnetic
field measurements at $z=0.48$ m for $L= 0.92$ m. The previously
determined radius $a=2 \pm 0.2$ cm is located at the edge
of the profile where $J_z(a)\approx0.05\times J_z(0)$.
Similar conclusions follow
from the magnetic diffusivity, derived from the measured electron
temperature profile and the observed instability growth time~$\tau_G$.

It should be emphasized here that the measurements in this
Letter only characterize the global structure of the external kink
mode. Although some of the observations
(e.g., axial structure and rotation frequency of the mode) show some resemblance
with measurements of internal drift-Alfv\'{e}n
waves in finite-$\beta$ plasmas with axial current\cite{Tang:1976},
additional internal measurements will be needed to
investigate the presence of these waves in our experiment.

In summary, we have presented first experimental results for the
current driven kink instability in a plasma column with one free end
that is not line-tied to its boundary due to a finite sheath
resistance. Mode structure and instability threshold are
significantly different from predictions from the Kruskal-Shafranov
theory and are accurately reproduced by a recently developed kink
theory. The finite rotation of the mode, which is also observed in
other laboratory devices\cite{Zuin:2004, Bergerson:2006}, is due to
the plasma flow along the helically kinked plasma column.

Though not addressed here, the effect of a non line-tied
end should be important for the relaxing kinked plasma column.
A destabilizing~$J_{\theta} \times B_z$~force coexists with
a stabilizing curvature force due to axial field line bending.
The free end allows shifts with reduced curvature and restoring force
possibly resulting in a helical saturated state of larger displacement than the line-tied case.
Numerical simulations are being implemented for the non-linear evolution of the kink mode.

Support by \textit{Los Alamos Laboratory Directed Research and
Development - Exploratory Research} program, and
\textit{Associazione Sviluppo Piemonte} for S. Abbate is gratefully
acknowledged.


\begin{thebibliography}{17}
\expandafter\ifx\csname
natexlab\endcsname\relax\def\natexlab#1{#1}\fi
\expandafter\ifx\csname bibnamefont\endcsname\relax
\def\bibnamefont#1{#1}\fi \expandafter\ifx\csname
bibfnamefont\endcsname\relax \def\bibfnamefont#1{#1}\fi
\expandafter\ifx\csname citenamefont\endcsname\relax
\def\citenamefont#1{#1}\fi \expandafter\ifx\csname
url\endcsname\relax \def\url#1{\texttt{#1}}\fi
\expandafter\ifx\csname
urlprefix\endcsname\relax\def\urlprefix{URL }\fi
\providecommand{\bibinfo}[2]{#2}
\providecommand{\eprint}[2][]{\url{#2}}
\bibitem[{\citenamefont{Kruskal}(1954)}]{Kruskal:1958}
\bibinfo{author}{\bibfnamefont{M.}~\bibnamefont{Kruskal}}
and \bibinfo{author}{\bibfnamefont{J.}~\bibnamefont{Tuck}},
\bibinfo{journal}{Proc. R. Soc. A} \textbf{\bibinfo{volume}{245}},
\bibinfo{pages}{222} (\bibinfo{year}{1958}).
\bibitem[{\citenamefont{Shafranov}(1956)}]{Shafranov:1956}
\bibinfo{author}{\bibfnamefont{V.~D.}~\bibnamefont{Shafranov}},
\bibinfo{journal}{At. Energ.} \textbf{\bibinfo{volume}{5}},
\bibinfo{pages}{38} (\bibinfo{year}{1956}).
\bibitem[{\citenamefont{Solov'ev}(1971)}]{Solove'ev:1971}
\bibinfo{author}{\bibfnamefont{L.~S.}~\bibnamefont{Solov'ev}},
\bibinfo{journal}{Sov. At. Energy} \textbf{\bibinfo{volume}{30}},
\bibinfo{pages}{14} (\bibinfo{year}{1971}).
\bibitem[{\citenamefont{Raadu}(1972)}]{Raadu:1972}
\bibinfo{author}{\bibfnamefont{M.~A.}~\bibnamefont{Raadu}},
\bibinfo{journal}{Sol. Phys.} \textbf{\bibinfo{volume}{22}},
\bibinfo{pages}{425} (\bibinfo{year}{1972}).
\bibitem[{\citenamefont{H. Baty}(1997)}]{Baty:1997}
\bibinfo{author}{\bibfnamefont{H.}~\bibnamefont{Baty}},
\bibinfo{journal}{Astron. Astrophys.} \textbf{\bibinfo{volume}{318}},
\bibinfo{pages}{621} (\bibinfo{year}{1997}).
\bibitem[{\citenamefont{A.W.~Hood}(1992)}]{Hood:1992}
\bibinfo{author}{\bibfnamefont{A.~W.}~\bibnamefont{Hood}},
\bibinfo{journal}{Plasma Phys. Contr. Fus.}
\textbf{\bibinfo{volume}{34}},
\bibinfo{pages}{411} (\bibinfo{year}{1992}).
\bibitem[{\citenamefont{D.L.~Meier}(2001)}]{Meier:2001}
\bibinfo{author}{\bibfnamefont{D.~L.}~\bibnamefont{Meier}},
\bibinfo{author}{\bibfnamefont{S.}~\bibnamefont{Koide}} and
\bibinfo{author}{\bibfnamefont{Y.}~\bibnamefont{Uchida}},
\bibinfo{journal}{Science} \textbf{\bibinfo{volume}{291}},
\bibinfo{pages}{84} (\bibinfo{year}{2001}).
\bibitem[{\citenamefont{I. Lanskii}(1990)}]{Lanskii:1990}
\bibinfo{author}{\bibfnamefont{I.~M.}~\bibnamefont{Lanskii}} and
\bibinfo{author}{\bibfnamefont{A.~I.}~\bibnamefont{Shchetnikov}},
\bibinfo{journal}{Sov. J. Plasma Phys.}
\textbf{\bibinfo{volume}{16}}, \bibinfo{pages}{322}
(\bibinfo{year}{1990}).
\bibitem[{\citenamefont{D.D.~Ryutov}(2004)}]{Ryutov:2004}
\bibinfo{author}{\bibfnamefont{D.~D.}~\bibnamefont{Ryutov}},
\bibinfo{author}{\bibfnamefont{R.~H.}~\bibnamefont{Cohen}} and
\bibinfo{author}{\bibfnamefont{L.~D.}~\bibnamefont{Pearlstein}},
\bibinfo{journal}{Phys. Plasmas} \textbf{\bibinfo{volume}{11}},
\bibinfo{pages}{4740} (\bibinfo{year}{2004}).
\bibitem[{\citenamefont{C.C. Hegna}(2004)}]{Hegna:2004}
\bibinfo{author}{\bibfnamefont{C.~C.}~\bibnamefont{Hegna}},
\bibinfo{journal}{Phys. Plasmas} \textbf{\bibinfo{volume}{11}},
\bibinfo{pages}{4230} (\bibinfo{year}{2004}).
\bibitem[{\citenamefont{Bergerson}(2006)}]{Bergerson:2006}
\bibinfo{author}{\bibfnamefont{W.~F.}~\bibnamefont{Bergerson}},
\bibinfo{author}{\bibnamefont{\emph{et~al.}}},
\bibinfo{journal}{Phys. Rev. Lett.} \textbf{\bibinfo{volume}{96}},
\bibinfo{pages}{015004} (\bibinfo{year}{2006}).
\bibitem[{\citenamefont{S.C.~Hsu}(2003)}]{Hsu:2004}
\bibinfo{author}{\bibfnamefont{S.~C.}~\bibnamefont{Hsu}} and
 \bibinfo{author}{\bibfnamefont{P.~M.}~\bibnamefont{Bellan}},
 \bibinfo{journal}{Phys.~Rev.~Lett.} \textbf{\bibinfo{volume}{90}},
 \bibinfo{pages}{215002} (\bibinfo{year}{2003}).
\bibitem[{\citenamefont{Zuin}(2004)}]{Zuin:2004}
\bibinfo{author}{\bibfnamefont{M.}~\bibnamefont{Zuin}},
\bibinfo{author}{\bibnamefont{\emph{et~al.}}},
\bibinfo{journal}{Phys. Rev. Lett.}
\textbf{\bibinfo{volume}{92}}, \bibinfo{pages}{225003}
(\bibinfo{year}{2004}).
\bibitem[{\citenamefont{D.D.~Ryutov}(2005)}]{Ryutov:2005}
\bibinfo{author}{\bibfnamefont{D.~D.}~\bibnamefont{Ryutov}},
\bibinfo{author}{\bibnamefont{\emph{et~al.}}},
\bibinfo{journal}{Phys. Plasmas}
\textbf{\bibinfo{volume}{13}}, \bibinfo{pages}{032105}
(\bibinfo{year}{2006}).
\bibitem[{\citenamefont{Furno}(2003)}]{Furno:2003}
\bibinfo{author}{\bibfnamefont{I.}~\bibnamefont{Furno}},
\bibinfo{author}{\bibnamefont{\emph{et~al.}}},
\bibinfo{journal}{Rev. Sci. Instrum.} \textbf{\bibinfo{volume}{4}},
\bibinfo{pages}{141} (\bibinfo{year}{2003}).
\bibitem[{\citenamefont{Hemsing et~al.}(2005)
\citenamefont{Hemsing, Furno, and Intrator}}]{Hemsing:2005}
\bibinfo{author}{\bibfnamefont{E.}~\bibnamefont{Hemsing}},
\bibinfo{author}{\bibfnamefont{I.}~\bibnamefont{Furno}} and
\bibinfo{author}{\bibfnamefont{T.}~\bibnamefont{Intrator}},
\bibinfo{journal}{IEEE Trans. Plasma Sci.} \textbf{\bibinfo{volume}{33}},
\bibinfo{pages}{448} (\bibinfo{year}{2005}).
\bibitem[{\citenamefont{Tang et~al.}(1976)
\citenamefont{Tang and Luhmann}}]{Tang:1976}
\bibinfo{author}{\bibfnamefont{J.~T.}~\bibnamefont{Tang}},
\bibinfo{author}{\bibnamefont{\emph{et~al.}}},
\bibinfo{journal}{Phys. Rev. Lett.}
\textbf{\bibinfo{volume}{34}},
\bibinfo{pages}{70} (\bibinfo{year}{1975});
\bibinfo{author}{\bibfnamefont{J.~T.}~\bibnamefont{Tang}} and
\bibinfo{author}{\bibfnamefont{N.~C.}~\bibnamefont{Luhman~Jr.}},
\bibinfo{journal}{Phys. Fluids}
\textbf{\bibinfo{volume}{19}},
\bibinfo{pages}{1935} (\bibinfo{year}{1976}).
\end{thebibliography}
\end{document}